\documentclass[%preprint,           % Format : preprint, twocolumn
               preprint,
               noshowpacs,          % Pacs : showpacs, noshowpacs
               nopreprintnumbers,     % Preprint: preprintnumbers,
               aps,                 % Society: 
               prd,                 % Journal Style : pra, prb, prc, prd, pre,
	       superscriptaddress,  % Affiliation (Title) : groupedaddress,
               nofootinbib,         % Footnote: footinbib, nofootinbib
               tightenlines,        % Remove additional spaces in a line
               floats,floatfix    % Floating pictures and tables
               ]{revtex4-1}

\usepackage{amsmath, amsfonts, amsthm, amssymb, graphicx}
\usepackage{color}
\usepackage[utf8]{inputenc}
\usepackage{lipsum}
\usepackage{centernot}
\usepackage{ulem}
\usepackage{comment}
\usepackage{feynmp}

\def\be{\begin{equation}}
\def\ee{\end{equation}}
\def\del{\partial}
\newcommand{\order}{{\cal O}}
 at 10truept
\newcommand{\citeseq}{\cite{KP1,KP2,KP3,KPSZ, KPS, KP4,etude}}

\DeclareMathAlphabet\mathbfcal{OMS}{cmsy}{b}{n}

\begin{document}

\title{A natural theory of dark energy}
\author{Lorenzo Bordin} 
\email{lorenzo.bordin@nottingham.ac.uk}
\affiliation{School of Physics and Astronomy, 
University of Nottingham, Nottingham NG7 2RD, UK} 
\author{Francesc Cunillera} 
\email{francesc.cunilleragarcia@nottingham.ac.uk}
\affiliation{School of Physics and Astronomy, 
University of Nottingham, Nottingham NG7 2RD, UK} 
\author{Antoine Leh\'ebel} 
\email{antoine.lehebel@nottingham.ac.uk}
\affiliation{School of Physics and Astronomy, 
University of Nottingham, Nottingham NG7 2RD, UK} 
\author{Antonio Padilla} 
\email{antonio.padilla@nottingham.ac.uk}
\affiliation{School of Physics and Astronomy, 
University of Nottingham, Nottingham NG7 2RD, UK}

\date{\today}

\begin{abstract}
We propose a mechanism that generates a naturally light dark energy field (with Hubble scale mass), starting from a theory with exclusively high scale (Planckian) couplings. It is derived from the clockwork model, with $\order(100)$ scalar fields interacting among themselves as well as with a 4-form field strength. We explicitly embed the key features of our model in type IIA supergravity. We also give an alternative interpretation in  a braneworld set-up.  
\end{abstract}
\maketitle

\section{Introduction}
The expansion of the universe is accelerating.  There is now strong evidence for this acceleration coming from  a slew of cosmological data, mostly notably the observation of distant supernova \cite{SN1,SN2} and measurements of the cosmic microwave background radiation \cite{CMB}.  This acceleration is feeble, driven by dark energy with an energy scale set by the Hubble constant, $H_0 \sim 10^{-33}$ eV \cite{derev}.  If dark energy corresponds to vacuum energy, its energy density is at least sixty orders of magnitude or more below its natural value  \cite{wein,pol,me,cliff}.  A popular alternative  is quintessence in which dark energy is driven by 
 a dynamical scalar field,  mirroring the inflationary dynamics of the early universe, albeit at a much lower scale.  However, if dark energy is truly dynamical, one has to assume that the large vacuum energy is cancelled using some other mechanism\footnote{An interesting proposal for eliminating vacuum energy is given by {\it vacuum energy sequestering} \citeseq. This set-up is particularly appealing because it eliminates vacuum energy whilst allowing for acceleration through a slowly rolling inflaton or  quintessence field \cite{KP2}.}, so that the current cosmic acceleration is  entirely due to the scalar dynamics.   Even with this compromise there is often a further tuning, since the effective mass of the scalar must lie at or below the Hubble scale in order to be cosmologically relevant on the largest scales today.  In a typical quintessence model, the  light mass  is usually put in by hand by assuming a very wide potential, but this is difficult to engineer from a consistent microscopic theory where the typical mass scales are much higher,  usually around the Planck scale.  Further, as is well known from the electroweak hierarchy problem, light scalar masses are susceptible to large radiative corrections \cite{Giudice}.

In this paper, we propose a model of dark energy driven by a pseudo-scalar field whose super-light mass emerges naturally from a simple microscopic theory with uniquely high scale couplings.  The model is a dark energy  avatar of the  pion, its low mass emerging from the spontaneous breaking of a weakly broken symmetry.  It consists of $\order(100)$ pseudo scalar fields (axions) with non-trivial mass mixing, one of which has a bilinear mixing with a 4-form field strength.  All mass scales in the theory are assumed to lie at, or close to, the Planck scale.   We will explicitly demonstrate how the key ingredients can arise naturally in string compactifications. 

The model is a marriage of two axion models that have been developed in recent years for different reasons. The first is the clockwork axion \cite{cw1, cw2} (see also \cite{GM1, Craig1, GM2, GM3, Teresi,Ahmed, Hambye, cwgravity, Torre, fimp, Agrawal, Im1, Im2, Martin} for related work and interesting applications), proposed in order to account for the super-Planckian decay constants required by models of cosmological relaxation \cite{relaxion}. The basic idea is to have a modest number of axions, $\phi_i$, whose masses mix with some characteristic strength $q>1$, so that they behave like the gears of a clockwork. If an external source is coupled with some strength to one end of the clockwork,  then the resulting low energy effective field theory (EFT) contains a zero mode that couples to the source with an exponentially weaker coupling.  The structure of the mass mixings are crucial. They allow for a nonlinearly realised shift symmetry  on the axions $\phi_i\to \phi_i +c q^{-i}$, where the size of the shift is warped by factors of $q$ as we move through the gears.   This symmetry can be  identified with an asymmetrically distributed unbroken subgroup of an underlying $U(1)^N$, with the axions identified as pions below the scale of spontaneous symmetry breaking.  Further, the shift symmetry guarantees the existence of a zero mode in the low energy EFT, while the warping ensures that its overlap with the far end of the clockwork is small. This is what suppresses the coupling to external sources. 

The second model in the marriage is the field theory model of axions coupled to 4-forms originally proposed by Dvali \cite{Dvali1,Dvali2}, then cleverly applied to cosmology by Kaloper and Sorbo \cite{KS1, KS2} (see also \cite{KLS}). This has been proposed as a field theory realisation of axion monodromy in string theory \cite{monod1,monod2}. The idea is to introduce a bilinear mixing between the axion and the 4-form field strength. The theory admits a dual description in terms of a massive pseudoscalar and the magnetic flux of the 4-form. The latter is locally constant in spacetime although it can jump between quantised values across a three dimensional membrane.   The model is of particular interest to single field inflation since it avoids some of the problems associated with super-Planckian field excursions for the inflaton.  This is because the effective inflaton is a gauge invariant combination of the axion and the magnetic flux - its large field values are obtained through the flux, which in turn may be identified with macroscopic quantities characterizing the system rather than high scale excitations of the inflaton field \cite{london}.  Small deformations of these models can also give rise to an emergent theory of vacuum energy sequestering, screening the effects of vacuum energy at large scales \cite{monap, monbasem, monnk}.

In the Dvali-Kaloper-Sorbo model, the inflaton gets a mass from its mixing with the 4-form field strength, and the stronger the coupling the larger the effective mass. It is this mechanism for generating a mass that we exploit in our model of dark energy. By coupling one end of our clockwork to the 4-form we guarantee that the mixing with the zero mode of the clockwork is exponentially small and as a result, a very small mass is generated in the low energy effective theory.  It should be relatively straightforward to generate the key ingredients of these  models from fundamental theory, and we illustrate this with a toy model involving a compactification of type IIA string theory on a K\"ahler manifold of the form $T^2 \times \Sigma_{g}^4$ where  $\Sigma_{g}^4$  is a   manifold of dimension four, and sufficiently high genus. To further enhance the prospect of deriving this theory as the low energy limit of some UV complete theory, we use the clockwork to deconstruct an extra dimension. This motivates a family of braneworld configurations where the low scale of dark energy emerges naturally on the brane. 

The rest of this paper is organised as follows: in the next section we build the discrete version of our model as a marriage of the two axion proposals described above. In section \ref{sec:cesc} we demonstrate how such a set-up can be obtained from a toy compactification of type IIA string theory. In section \ref{sec:dec} we perform the dimensional deconstruction of the discrete model, and study a family of braneworld configurations showing how the braneworld obsverver measures a low scale of dark energy on account of the warping of the extra dimension. In section \ref{sec:conc}, we conclude. 

\section{Clockwork dark energy} 
\label{sec:discrete}

We begin with a chain of $N+1$ pseudo-scalar fields $\phi_0$, ..., $\phi_N$, all characterized by a single  ultraviolet mass scale $M$ and a nearest neighbour interaction with strength $q$  \cite{cw1, cw2}.  The mass $M$ is assumed to lie at or close to the Planck scale. We further assume that one end of the chain is coupled to a 4-form field strength as in the Dvali-Kaloper-Sorbo model. The combined set-up is described by the following  Lagrangian density:
%commentstart
\begin{multline}
\mathcal{L}=-\dfrac12\left[\sum_{i=0}^N(\partial\phi_i)^2+M^2\sum_{i=0}^{N-1}(\phi_i-q\phi_{i+1})^2\right]+ \dfrac{\mu}{24}\phi_N \frac{\epsilon^{\mu\nu\alpha\beta} }{\sqrt{-g}}F_{\mu\nu\alpha\beta}  -\dfrac{1}{48} F_{\mu\nu\alpha\beta} F^{\mu\nu\alpha\beta} \ .
\label{eq:L0}
\end{multline}
%commentend

In the first line we recognise the clockwork model of \cite{cw1, cw2}.  In principle we could allow for site dependent masses $M_i \sim M$ and mixing strengths $q_i \sim q$, although for simplicity we take them all to be equal.  The coupling $q$ is dimensionless and assumed to be greater than 1, but it remains of order unity. The second line of \eqref{eq:L0} contains the Dvali-Kaloper-Sorbo model for the $N$th site in the chain.  Here $F_{\mu\nu\rho\sigma}=4 \del_{[\mu} A_{\nu\alpha\beta]}$ is the 4-form field strength, $\epsilon^{\mu\nu\rho\sigma}$ is the totally antisymmetric Levi-Civita symbol defined such that $\epsilon^{0123}=1$ and indices are raised and lowered with respect to the metric $g_{\mu\nu}$.  The bilinear mixing between the axion and the 4-form reveals another mass scale, $\mu \sim M$, which we also assume to be given by the characteristic ultraviolet scale of the theory. The gravitational sector of the theory is assumed to be described by Einstein gravity, although we will not need to include that in our discussion. 

It is well-known that the clockwork Lagrangian gives rise to a massless Goldstone pseudo-scalar \cite{cw1,cw2,GM1}, a consequence of the non-linearly realised  shift  symmetry $\phi_i \to \phi_i + cq^{-i}$ for arbitrary $c$. This symmetry remains perturbatively unbroken through the mixing with the 4-form, although in the presence of charged membranes, non-perturbative effects break the continuous symmetry down to a discrete subgroup \cite{Pol1, BP}. However, the symmetry is only {\it mildly} broken because $\phi_N$ has exponentially suppressed overlap with the zero mode. We thus expect that the zero mode acquires a mass, but that the latter remains tiny. In order to see this explicitly  it is convenient  to integrate out the 4-form and pass to a dual description in which the 4-form mixing generates a new mass term for the last axion in the chain.  This can be done in a straightforward manner by adding a Lagrange multiplier term of the form $\frac{1}{24}Q \frac{\epsilon^{\mu\nu\alpha\beta}}{\sqrt{-g}} (F_{\mu\nu\rho\sigma}-4\del_{[\mu}A_{\nu\rho\sigma]} )$ then eliminating $F$ using its algebraic equation of motion\footnote{Note that since the theory is quadratic in $F$ this amounts to performing the Gaussian  in the path integral, which can,  of course, be done exactly.} \cite{KS1}.  This yields a  dual theory described by the following Lagrangian:
%commentstart
\begin{multline}
\mathcal{L}=-\dfrac12\left[\sum_{i=0}^N(\partial\phi_i)^2+M^2\sum_{i=0}^{N-1}(\phi_i-q\phi_{i+1})^2\right]-\dfrac12 (Q+\mu \phi_N)^2- \frac{Q}{6} \frac{\epsilon^{\mu\nu\alpha\beta} }{\sqrt{-g}}\del_{[\mu}A_{\nu\alpha\beta]}  \ .
\label{eq:Ldual}
\end{multline}
%commentend
The Lagrange multiplier $Q$ is fixed to be constant by the variation of the 3-form.  If the latter is coupled to membrane charges, then $Q$ is quantised in units of the membrane charge, $e$, as in  $\langle Q\rangle=2\pi \mathcal{N}_Q e$ for integer values of $\mathcal{N}_Q$ \cite{Pol1, BP}.  This quantisation condition is compatible with the unbroken symmetry transformations, which take the form
$
\phi_i\to \phi_i+2\pi n \frac{e}{\mu}q^{N-i}, \ Q \to Q+2\pi ne
$
for integer values of  $n$. The mass matrix in the dual description is given by
%commentstart
{\small
\begin{equation}
{\cal M}_{ij}=M^2\begin{bmatrix}
1 & -q & 0 & & \cdots & & 0 \\
-q & 1+q^2 & -q & & & & \\
0 & -q & 1+q^2 & \ddots & \ddots & & \vdots \\
 & & \ddots & \ddots & \ddots & & \\
\vdots & & \ddots & \ddots & 1+q^2 & -q & 0 \\
 & & & & -q & 1+q^2 & -q \\
0 & & \cdots & & 0 & -q & r+q^2
\label{Eq:matrixG}
\end{bmatrix},
\end{equation}}
%commentend
where $r=(\mu/M)^2$ is the square of the ratio between the two mass scales. The eigenmasses of this matrix are given by  the roots of an $(N+1)$th order polynomial. It turns out that there is a tower of $N$ massive modes whose masses go with the ultraviolet scale $M$.  The remaining mode is massless in the limit where $r \to 0$, and in general is ultralight. We can find it by linearising the above eigenvalue problem; the resulting ultralight mass scale is given by
%commentstart
\begin{equation}
\begin{split}
m_0^2&\simeq\frac{(q^2-1)^2 r M^2}{q^{2 (N+1)}(q^2+r -1)+(N+1) r(1-q^2) -q^2-r +1} .
\label{eq:m0}
\end{split}
\end{equation}
After integrating out the four-form and setting the corresponding flux to zero, the emergent  potential for the ultra-light mode, $\pi_0$, is given by a simple quadratic 
\be
V_0= \frac12 m_0^2 \pi_0^2
\ee
%commentend
In the limit of large $N$ (and for $q$ larger than unity), $m_0$ is well approximated by
%commentstart
\begin{equation}
m_0^2\simeq\dfrac{1}{q^{2 (N+1)}}\dfrac{(q^2-1)^2 r M^2}{(q^2+r -1)}.
\label{eq:discrm0largeN}
\end{equation}
%commentend
Thus, $m_0$ acquires an exponential suppression with respect to $M$, the argument in the exponential being the total number of clockwork gears, $N+1$. If we take $M=\mu=m_\mathrm{P}$ ($m_\mathrm{P}=1.22\times10^{19}$~GeV being the Planck mass), $q=2$ and $N=200$, we get $m_0\simeq5.7\times10^{-33}$~eV,  which is the energy scale associated with the Hubble rate.

At large scales, the dynamics will be equivalent to quintessence driven by a quadratic potential. However, the mass scale of  potential has not been tuned to the tiny value demanded by nature, rather it has arisen naturally on account of the clockwork mechanism and the coupling to the 4-form.  The underlying theory is made up uniquely of high scale couplings. We emphasize the fact that the mass mixings need not be identical but can have site dependence. As long as they are  greater than unity, the clockwork mechanism will kick in as usual, even for order one couplings, and the suppression of the mass of the ultralight mode will occur as desired. 

Finally we note that if we allowed for a non-vanishing flux, the potential would actually go as 
\be
V_0=\frac12m_0^2  \left(\frac{2 \pi \mathcal{N}_Q e}{m_0}+ \pi_0\right)^2
\ee
For natural values of $\pi_0 \sim m_P$, the flux term dominates the potential  for non-vanishing $\mathcal{N}_Q$ and there is too much dark energy. This is why we assumed a vanishing flux $\mathcal{N}_Q =0$, which is a robust condition provided the nucleation rate for bubbles of non-vanishing flux is suppressed\footnote{Although a detailed analysis of this interesting question is beyond the scope of our work, we  expect that there is indeed suppression for the following reason. For $e \sim m_P^2$ a unit of flux generates a large positive Planckian potential. Therefore, we can crudely model the transition from vanishing to non-vanishing flux in terms of a bubble of Planckian de Sitter curvature nucleating in a quasi de Sitter background with a very small curvature.  Such processes, tunnelling from true to false, are possible but are known to be suppressed relative to those going from false to true, becoming infinitely suppressed in the limit that the low scale curvature approaches zero (see e.g. \cite{dsds}) .}.

\section{Clockwork dark energy from string theory} \label{sec:cesc}
Our model can be motivated from the point of view of a higher dimensional theory. Indeed, in a theory with extra dimensions, a large number of scalar fields in the 4D EFT is often associated with the periods of differential p-forms living in the higher dimensional theory. In \cite{Valenzuela,Marchesano}, the authors showed that one may rewrite the democratic type IIA supergravity formulation \cite{democratic} in terms of a pseudo-action containing Minkowski 4-forms and its dual fields, which is equivalent to the democratic action at the level of the equations of motion.

The democratic action is usually written in terms of a polyform gauge invariant field strength, $G=G_0+G_2+\ldots +G_{10}$ where  $\mathbf{G}=d\mathbf{C}+B\wedge\mathbf{C}+\mathbfcal{F}\wedge e^B$  and $B$ is the Kalb-Ramond 2-form,  $\mathbf{C}=C_1+C_3+\ldots +C_9$ the polyform gauge field and $\mathbfcal{F}= \mathcal{F}_0+\mathcal{F}_2+\ldots +\mathcal{F}_{10}$ a formal sum of internal fluxes only. Let us take the following ansatz for the non-vanishing supergravity fields
 \begin{gather}
 B_2 = b^i(x) \omega_i, \quad C_3= c_3^0(x), \quad C_5=c_3^i(x)\omega_i\ , \nonumber\\
 C_7=\bar{c}_{3i}(x)\bar{\omega}^i, \quad \mathcal{F}_2=q^i \omega_i,\quad \mathcal{F}_4=e_i\bar{\omega}^i,
 \end{gather}
where we note that this corresponds to the massless limit of type IIA supergravity.  We have introduced the cohomology basis of \{2,4\}-forms in the internal manifold $\mathcal{M}_6$ as \{$\omega_i$, $\bar{\omega}^i$\}, respectively, and $\omega_6$ will denote the volume form of $\mathcal{M}_6$. With this ansatz, $\mathbf{G}$ reads\footnote{We have chosen to use the compact notation
\begin{gather}
F_4^0=dc_3^0, \quad F_4^i=dc_3^i+b^i F_4^0,\quad \bar{F}_{4i}=d\bar{c}_{3i}+\mathcal{K}_{ijk}b^j dc_3^k. \nonumber
\end{gather}} 
%commentstart
\begin{gather}
G_2=q^i \omega_i,\qquad G_4=F_4^0+\left(e_i+\mathcal{K}_{ijk}b^jq^k\right)\bar{\omega}^i, \nonumber\\
G_6=F_4^i\omega_i+\left(e_i b^i+\frac{1}{2}\mathcal{K}_{ijk}b^ib^jq^k\right)\omega_6,\quad G_8=\bar{F}_{4i}\bar\omega^i,
\end{gather}
%commentend
with $\mathcal{K}_{ijk}=\int_{\mathcal{M}_6}\omega_i\wedge\omega_j\wedge\omega_k$, the triple intersection numbers of $\mathcal{M}_6$.

Below we describe the effective 4D action, in doing so we will assume that the volume moduli, including the dilaton, are stabilised by some other ingredient of the theory (see e.g \cite{Saltman}). Our main objective is to obtain the main characteristic of the clockwork mechanism after compactifying the theory down to four dimensions. 

These fields enter the 4D potential as follows (see for details \cite{Valenzuela,Marchesano})
%commentstart
\begin{gather}
V_{4D}\propto F_4^0\wedge* F_4^0 + g_{ij}F_4^i\wedge* F_4^j+{g}^{ij}\bar{F}_{4i}\wedge* \bar{F}_{4j}+ F_4^0\rho_0 + F_4^i\rho_i + \bar{F}_{4i}\bar{\rho}^i\ , 
\end{gather}
%commentend
where $g_{ij}=\int_{\mathcal{M}_6}\omega_i\wedge\star\omega_j$, $g^{ij}=\int_{\mathcal{M}_6}\bar{\omega}^i\wedge\star\bar{\omega}^j$ and we have introduced the dual scalars to the Minkowski 4-forms
%commentstart
\begin{gather}
l_s \rho_0=e_i b^i + \frac{1}{2} \mathcal{K}_{ijk}q^ib^jb^k, \nonumber \\
-l_s \rho_i= e_i + \mathcal{K}_{ijk}q^jb^k, \quad l_s \bar{\rho}^i= q^i\ . \nonumber
\end{gather}
%commentend
Eliminating $F_4^i$ and $\bar{F}_{4i}$ through their equations of motion in favour of their dual scalars, the 4D Lagrangian density with only one Minkowski 4-form is given by
%commentstart
\begin{multline}
\hspace{-0.5cm}\mathcal{L}_{4D}=-\frac{1}{\kappa_4^2}\left[\dfrac{e^{-2\phi}}{4} g_{ij}db^i\wedge * db^j+\frac{\hat{V}^3_6 e^{\frac{\phi}{2}}}{32}F_4^0\wedge* F_4^0  - \frac{1}{4}F_4^0\rho_0 + \frac{l_s^6 e^{\frac{\phi}{2}}}{2 \hat{V}^2_6 }\left(g^{ij}\rho_i\rho_j+e^\phi g_{ij}\bar{\rho}^i\bar{\rho}^j\right)\right]\ ,
\label{eq:L4D}
\end{multline}
%commentend
with $\text{Vol}\left(\mathcal{M}_6\right)=l_s^6\hat{V}_6$.

As this stands, the vacuum expectation value for the axions is not necessarily vanishing, owing to the presence of a tadpole in the potential for $b_i$.  We can fix this by shifting $b_i \to \beta_i+ b_i$ for some  constant flux $\beta_i$ chosen so that it eliminates the tadpole. By comparison to \eqref{eq:L0}, one finds a Lagrangian density of the form\footnote{In writing \eqref{eq:4D_L} we have dropped  a next-to-leading order correction term to the coupling ($\sim F_4^0b^2$)  as well as a cosmological constant term  and a total derivative that do not contribute to our discussion.}
 \begin{gather}
\mathcal{L}=-\frac12 \left[\gamma_{ij} \del b^i \del b^j+M_{ij}b^i b^j +\dfrac{\alpha}{48} \left({F}_4^0\right)^2\right]+\frac{1}{24} \chi_i b^i F_4^0  \ ,
\label{eq:4D_L}
 \end{gather}
with
\be
\gamma_{ij} =\frac{e^{-2\phi}}{2 \kappa_4^2}\int_{\mathcal{M}_6}\omega_i\wedge\star\omega_j \ ,\quad \alpha=\frac{2\hat{V}_6^3 e^{\frac{\phi}{2}}}{3\kappa_4^2} \nonumber \ ,
\ee
\begin{equation}
M_{ij} =  \frac{e^{\frac{\phi}{2}}l_s^4}{\hat{V}^2_6 \kappa_4^2}  \mathcal{K}_{ikl}\mathcal{K}_{jmn} q^k q^m\int_{\mathcal{M}_6}\bar{\omega}^l\wedge\star\bar{\omega}^n\ , \quad\chi_i =\frac{6 }{\kappa_4^2 l_s} \left( e_i+\mathcal{K}_{ijk} \beta^j q^k\right)\ \nonumber
\end{equation}

The main ingredients of the clockwork dark energy model that we have considered so far are: a mass matrix with vanishing eigenvalue and a zero mode that overlaps with the remaining four form. The former is the clockwork condition while the latter allows the zero mode to acquire a small mass. Note that our goal here is to motivative these two key ingredients, rather than the precise form of the clockwork Lagrangian presented in the last section.  In any event, if $s^i$ is an eigenvector of $M_{ij}$ with null eigenvalue, then the first condition requires $M_{ij}s^i\sim\mathcal{K}_{ikl}q^ks^i=0$ while the second gives $s^i \chi_i \neq 0$. To comply with these two requisites, we must therefore impose the following condition on the internal geometry
\be
\int_{\mathcal{M}_6}\left(\omega_i s^ i\right)\wedge\omega_j\wedge\left(\omega_k q^k\right) = 0\ .
\label{eq:condition}
\ee
A simple solution to the above constraint is provided by $s^i=q^i$ and the corresponding 2-form given by a product of 1-forms, i.e. $\omega_kq^k=u \wedge v$. In that case, the constraint is immediately satisfied due to the antisymmetry of the wedge product on odd forms. As long as we require $q^i e_i \neq 0$, the second condition can also be satisfied.  

The proposed solution could be realised on a Calabi-Yau three-fold of sufficiently high genus.  However, a simplified way to visualise the right configuration is given in the presence of  a {\it K\"ahler} internal manifold of the form $\mathcal{M}_6\simeq T^2\times\Sigma_g^4$, where $g$ indicates the genus of the two-dimensional complex surface. Indeed, with this setup one naturally selects the zero mode, $q^i$, to point along the toroidal directions $T^2$. This is because the top form along $T^2$ is automatically a 2-form corresponding to  a product of the 1-forms of the torus.  As long as we take the $e_i$ to overlap slightly with the $T^2$, but not completely,  we also satisfy the second condition, allowing the zero mode to gain a small mass through its 4-form mixing. Note that an explicit solution to the Einstein equations for a generic  K\"ahler internal manifold that is sufficiently small and compact to be compatible with standard four dimensional observations may be difficult to find in practice and we present it here just for illustration. We leave a more detailed investigation to future study.

Finally, we have shown that for the clockwork mechanism to bring the mass of the lightest mode down to the dark energy scale one requires $\mathcal{O}\left(200\right)$ scalar fields. For simplicity of exposition, let us think of these fields as descending only from the Kalb-Ramond 2-form. Then, the previous requirement translates into a constraint on the cohomology structure of $\Sigma_g^4$ after fixing the $T^2$ geometry, namely $g=b_1-b_2/2$ where $b_p$ is the {\it p-th} Betti number of $\Sigma_g^4$. Using, $b_2\sim\mathcal{O}\left(200\right)$ we find that $g\sim b_1-100$.

\section{Deconstructing clockwork dark energy} \label{sec:dec}

If the number of pseudo-scalar fields is very high, $N\to \infty$, the clockwork mechanism can arise as a deconstruction of an extra compact dimension \cite{GM1}. In this limit, the clockwork gears merge into a single field $\Phi$, and the gear $\phi_i$ becomes the value of $\Phi$ at site $i$; the interaction of the 4-form with the last site of the discrete clockwork corresponds to the localisation of the 4-form on a brane at the boundary of the compactified extra dimension. Our model can be obtained from the following 5-D theory, defined on a fixed geometrical background, 
\be \label{eq:S05D}
S_{5\mathrm{D}}=S_\text{bulk} +S_0+S_{\pi R}
\ee
where we have a canonical scalar in the bulk 
\be
S_\text{bulk} =\int d^4x \int_0^{\pi R}dy \  \sqrt{-g} \left[-\dfrac{m}{2}\ g^{IJ} (\partial_I \Phi) (\partial_J \Phi) \right]
\ee
and two branes, one containing the dark energy sector,
\be
S_0=
\int_{y=0} d^4 x \sqrt{-\gamma_0}\Big(\dfrac{\mu}{24}\Phi \frac{\epsilon^{\mu\nu\alpha\beta} }{\sqrt{-\gamma_0} }F_{\mu\nu\alpha\beta} -\dfrac{1}{48} F^{\mu\nu\alpha\beta}F_{\mu\nu\alpha\beta}\Big)\Big]
\ee
and the other containing the matter sector
\be
S_{\pi R }=\int_{y={\pi R}} d^4 x \sqrt{-\gamma_R} \mathcal{L}_m (\gamma_R^{\mu\nu}, \Psi)
\ee
We define $g_{IJ}$  to be the 5-D metric, with indices $I, J$ running over the four spacetime dimensions $x^\mu$, as well as the additional fifth dimension $y$.  For its part, the $y$ dimension extends from 0 to $\pi R$, with a reflection symmetry at $y=0$ and $y=\pi R$. These two boundary surfaces represent the location of the branes with the induced metric on the brane at $y=0, \pi R$ given by $\gamma^0_{\mu\nu}, ~\gamma^R_{\mu\nu}$ respectively.  Upon dividing the $y$ dimension into $N+1$ sites and discretising the action \eqref{eq:S05D} accordingly,
one can recover the Lagrangian density \eqref{eq:L0}, provided we input the following geometry:
%commentstart
\be \label{eq:clockwork_geometry}
ds^2=e^\frac{4ky}{3} (dy^2+dx^2),
\ee
%commentend
where $dx^2$ entails the 4-D metric on a brane, and identifying $q^N=\mathrm{e}^{\pi kR}$ with the mass scale $M = N/(\pi R)$.  We imagine all such scales - $M, m, \mu$ and  $k$ - to be of the same order, corresponding to the UV scale of the underlying braneworld theory.  For this particular geometry, the latter corresponds to a five dimensional linear dilaton model with boundary terms living on the two branes \cite{GM2}.

However, the mechanism that suppresses the mass of the lightest mode is very general and works also for metrics that differ from \eqref{eq:clockwork_geometry}. 
For instance, it works also in the following family of metrics:
%commentstart
\be
ds^2 = e^{\frac{4ky}{3}} \left(e^{-4 \ell ky} \, dy^2 + dx^2 \right) \,.
\label{gen_metric}
\ee
%commentend
The clockwork geometry is recovered by choosing $\ell = 0$, while for $\ell = 1/3$ one gets Randall-Sundrum \cite{Randall:1999ee}. 
The equation of motion for $\Phi$, once $F$ has been integrated out, is
%commentstart
\be \label{eq:eqPhi}
e^{2(\ell-\frac13)ky} \left[\Phi'' + 2( 1+ \ell) k \Phi' + \frac{\Box  \Phi}{e^{4k\ell y}} \right] = \delta(y) \frac{\mu(\mu \Phi + Q)} {m},
\ee
%commentend
where a prime stands for a $y$ derivative, and $\Box$ is the 4-D d'Alembertian. Setting $\Phi=-Q/\mu+\delta\Phi$, we can easily solve for $\delta\Phi$ in the bulk by taking the 4-D Fourier transform of eq.~(\ref{eq:eqPhi})\footnote{In general, one should embed the action \eqref{eq:S05D} in some geometric theory and study the perturbations of the geometric quantities as well as the fluctuations of the clockwork field. However, these fluctuations decouple at linear level. We can thus consistently examine the fluctuations of $\Phi$ on their own.}.
 The solution is then given by
 %commentstart
\be \label{eq:solution_general_metric}
\begin{split}
\delta\Phi(y, x^\mu) &= \int d^4p \, \Big[ A(p^2) \ J_{\frac{1+\ell}{2\ell}} \Big(\frac{\sqrt{-p^2}}{2k\ell e^{2k\ell y}}\Big)+B(p^2) \ J_{-\frac{1+\ell}{2\ell}} \Big(\frac{\sqrt{-p^2}}{2k\ell e^{2k\ell y}}\Big) \Big]e^{i p_\mu x^\mu - k(1+\ell)y} ,
\end{split}
\ee
%commentend
where the $J_\alpha(z)$ are Bessel functions of the first kind, $A$ and $B$ being free functions. Equation \eqref{eq:eqPhi} further imposes some boundary conditions at $y=0$ and $\pi R$. As a result, $A$ and $B$ are linearly related, while $p^2$ can only take some values among a quantised set, $p^2=-m_n^2$, $n\in\mathbb{N}$. The masses $m_n$ are found as solutions of the following equation:
%commentstart
\begin{gather}
J_{\frac{\ell-1}{2 \ell}}\left(\frac{m_n}{2 k \ell}e^{-2 k \ell \pi  R}\right)\left[J_{\frac{\ell+1}{2 \ell}}\left(\frac{m_n}{2 k \ell}\right) \left[4 k (\ell+1)m+\mu ^2\right]-2 m_n m J_{\frac{3\ell+1}{2\ell}}\left(\frac{m_n}{2 k \ell}\right)\right]\nonumber \\
+J_{\frac{1-\ell}{2\ell}}\left(\frac{m_n}{2 k \ell}e^{-2 k \ell \pi  R}\right)\left[\mu ^2 J_{-\frac{\ell+1}{2 \ell}}\left(\frac{m_n}{2 k \ell}\right)-2 m_n m J_{\frac{\ell-1}{2 \ell}}\left(\frac{m_n}{2 k \ell}\right)\right]=0\ .
\end{gather}
%commentend
They are all of order $k$, except for the first which is very light, being suppressed by $e^{k \pi R} \gg 1$. It is possible to find an approximate expression for this light mass:
%commentstart
\be\label{eq:lightest_mode}
m_0^2 (\ell) \simeq \frac{\mu^2}{m} \frac{ {8 k^2 \ell (1-\ell^2)}} 
{2e^{2k(1-\ell)\pi R}\ell\left[2k(1+\ell)+\frac{\mu^2}{m}\right]-(1+\ell)\left(4k\ell +\frac{\mu^2}{m}\right)-e^{-4k\ell\pi R}(\ell-1) \frac{\mu^2}{m}}\ .
\ee
%commentend
The above approximation is based on an expansion of the Bessel functions assuming $m_0\ll k\ell$. This is no longer true when $\ell \to 0$. To compute the lightest mass in the clockwork geometry, one can solve directly 
$$m_0^2 (0) =\frac{4 k^2 \mu ^2}{e^{2 \pi  k R} \left(4 k m+\mu ^2\right)-4 k m-2 \pi  k \mu ^2 R+\mu ^2} \ .$$
\newline
Equation \eqref{eq:lightest_mode} shows that the suppression mechanism works for the whole family of metrics \eqref{gen_metric}, even if it gets less efficient as we depart from the clockwork geometry. As an example for the values given below Eq.~\eqref{eq:discrm0largeN}, and setting $\mu^2/m=M$, one gets $m_0(\ell=1/3)\simeq 6.6\times10^{-16}\ \text{eV}$ (Randall-Sundrum geometry), while $m_0(\ell= 0)\simeq 6.3\times10^{-33}\ \text{eV}$. Note that Eq.~\eqref{eq:lightest_mode} should not be trusted physically when $l\gtrsim1$, because the length of the $y$ dimension drops below the UV scale of the five dimensional theory.

As it stands, the mass of the lightest dark energy mode is suppressed relative to UV scales set by matter resident on {\it either} of the two branes. Why, then, have we placed matter on the brane at  $y=\pi R$?   This is because the energy density of the dark energy field in slow roll is enhanced by the effective four dimensional Planck scale. This enhancement exactly compensates for the suppression in the mass of the field, and  in the end the energy density during slow roll scales like $k^4$. We shall postpone further details to a future publication \cite{future}.  In any event, it turns out that the energy density of dark energy will  only be suppressed if we calibrate our scales relative to the $\pi R$ brane, where the warp factor is exponentially large. This is why we put the visible sector of our theory on the right hand brane. 

\section{Conclusions} \label{sec:conc}

In this paper, we have considered the interplay between clockwork gravity and the coupling between its pseudo-scalar fields and Minkowski top-forms. The result of this added coupling is a natural realisation of super-light dark energy whose energy scale arises solely from the characteristics of the high energy theory. 

Furthermore, we have considered two possible UV toy models for the EFT. Firstly, we consider a type IIA compactification. Upon constraining the geometry of the internal space, we find a solution that ensures that our toy models will contain a zero eigenvalue on its mass matrix, the key feature of the clockwork mechanism. Secondly, we propose a generalisation of \cite{GM1} which contains the desired coupling to a single 4-form living on a brane.  

It would be interesting to explore the phenomenology of our natural dark energy scenario in greater detail in each of its different guises.   In two particularly exciting developments, work is currently underway to develop the  toy supergravity set-up to tackle the coincidence problem, or to ease the tension between measurements of the current Hubble scale. 

%commentstart
\begin{acknowledgments}
\vskip.5cm

{\bf Acknowledgments}: 
  A.P. is  funded by a Leverhulme Trust
  Research Project Grant. L.B., A.L. and A.P. are funded by an STFC Consolidated Grant. F.C. is funded by a studentship at the University of Nottingham.
\end{acknowledgments}
%commentend


\begin{thebibliography}{99}


%\cite{Perlmutter:1998np}
\bibitem{SN1}
 %\cite{Perlmutter:1998np}
  S.~Perlmutter {\it et al.} [Supernova Cosmology Project Collaboration],
  %``Measurements of $\Omega$ and $\Lambda$ from 42 high redshift supernovae,''
  Astrophys.\ J.\  {\bf 517} (1999) 565
%  doi:10.1086/307221
  [astro-ph/9812133].
  %%CITATION = doi:10.1086/307221;%%
  %12078 citations counted in INSPIRE as of 05 Dec 2019
\bibitem{SN2}
  A.~G.~Riess {\it et al.} [Supernova Search Team],
  %``Observational evidence from supernovae for an accelerating universe and a cosmological constant,''
  Astron.\ J.\  {\bf 116} (1998) 1009
%  doi:10.1086/300499
  [astro-ph/9805201].
  %%CITATION = doi:10.1086/300499;%%
  %11582 citations counted in INSPIRE as of 24 Jul 2019
%\cite{Riess:2019cxk}

%\cite{Aghanim:2018eyx}
\bibitem{CMB}
  N.~Aghanim {\it et al.} [Planck Collaboration],
  %``Planck 2018 results. VI. Cosmological parameters,''
  arXiv:1807.06209 [astro-ph.CO].
  %%CITATION = ARXIV:1807.06209;%%
  %1183 citations counted in INSPIRE as of 24 Jul 2019
  
  %\cite{Copeland:2006wr}
\bibitem{derev}
  E.~J.~Copeland, M.~Sami and S.~Tsujikawa,
  %``Dynamics of dark energy,''
  Int.\ J.\ Mod.\ Phys.\ D {\bf 15} (2006) 1753
 % doi:10.1142/S021827180600942X
  [hep-th/0603057].
  %%CITATION = doi:10.1142/S021827180600942X;%%
  %3887 citations counted in INSPIRE as of 24 Jul 2019
  
\bibitem{wein}
%\cite{Weinberg:1987dv}
%\bibitem{Weinberg:1987dv} 
S.~Weinberg,
%``The Cosmological Constant Problem,''
Rev.\ Mod.\ Phys.\  {\bf 61}, 1 (1989).
%%CITATION = RMPHA,61,1;%%

%\cite{Polchinski:2006gy}
\bibitem{pol} 
  J.~Polchinski,
  %``The Cosmological Constant and the String Landscape,''
  hep-th/0603249.
  %%CITATION = HEP-TH/0603249;%%
  %155 citations counted in INSPIRE as of 17 Mar 2015

%\cite{Burgess:2013ara}
\bibitem{cliff} 
  C.~P.~Burgess,
  %``The Cosmological Constant Problem: Why it's hard to get Dark Energy from Micro-physics,''
  arXiv:1309.4133 [hep-th].
  %%CITATION = ARXIV:1309.4133;%%
  %19 citations counted in INSPIRE as of 17 Mar 2015

%\cite{Padilla:2015aaa}
\bibitem{me}
  A.~Padilla,
  %``Lectures on the Cosmological Constant Problem,''
  arXiv:1502.05296 [hep-th].
  %%CITATION = ARXIV:1502.05296;%%
  
    %\cite{Giudice:2008bi}
\bibitem{Giudice}
  G.~F.~Giudice,
  %``Naturally Speaking: The Naturalness Criterion and Physics at the LHC,''
  In *Kane, Gordon (ed.), Pierce, Aaron (ed.): Perspectives on LHC physics* 155-178
%  doi:10.1142/9789812779762_0010
  [arXiv:0801.2562 [hep-ph]].
  %%CITATION = doi:10.1142/9789812779762_0010;%%
  %115 citations counted in INSPIRE as of 11 Jun 2018
  
  %\cite{Choi:2015fiu}
\bibitem{cw1}
  K.~Choi and S.~H.~Im,
  %``Realizing the relaxion from multiple axions and its UV completion with high scale supersymmetry,''
  JHEP {\bf 1601} (2016) 149
 % doi:10.1007/JHEP01(2016)149
  [arXiv:1511.00132 [hep-ph]].
  %%CITATION = doi:10.1007/JHEP01(2016)149;%%
  %138 citations counted in INSPIRE as of 01 Aug 2019

  %\cite{Kaplan:2015fuy}
\bibitem{cw2}
  D.~E.~Kaplan and R.~Rattazzi,
  %``Large field excursions and approximate discrete symmetries from a clockwork axion,''
  Phys.\ Rev.\ D {\bf 93} (2016) no.8,  085007
%  doi:10.1103/PhysRevD.93.085007
  [arXiv:1511.01827 [hep-ph]].
  %%CITATION = doi:10.1103/PhysRevD.93.085007;%%
  %151 citations counted in INSPIRE as of 01 Aug 2019
  
  %\cite{Giudice:2016yja}
\bibitem{GM1}
  G.~F.~Giudice and M.~McCullough,
  %``A Clockwork Theory,''
  JHEP {\bf 1702} (2017) 036
%  doi:10.1007/JHEP02(2017)036
  [arXiv:1610.07962 [hep-ph]].
  %%CITATION = doi:10.1007/JHEP02(2017)036;%%
  %98 citations counted in INSPIRE as of 01 Aug 2019
  %\cite{Craig:2017cda}

\bibitem{Valenzuela}
  S.~Bielleman, L.~E.~Ibanez and I.~Valenzuela,
  %``Minkowski 3-forms, Flux String Vacua, Axion Stability and Naturalness,''
  JHEP {\bf 1512} (2015) 119
%  doi:10.1007/JHEP12(2015)119
  [arXiv:1507.06793 [hep-th]].
  %%CITATION = doi:10.1007/JHEP12(2015)119;%%
  %48 citations counted in INSPIRE as of 01 Dec 2019

\bibitem{Marchesano}
  F.~Carta, F.~Marchesano, W.~Staessens and G.~Zoccarato,
  %``Open string multi-branched and Kähler potentials,''
  JHEP {\bf 1609} (2016) 062
%  doi:10.1007/JHEP09(2016)062
  [arXiv:1606.00508 [hep-th]].
  %%CITATION = doi:10.1007/JHEP09(2016)062;%%
  %43 citations counted in INSPIRE as of 01 Dec 2019

\bibitem{democratic}
  E.~Bergshoeff, R.~Kallosh, T.~Ortin, D.~Roest and A.~Van Proeyen,
  %``New formulations of D = 10 supersymmetry and D8 - O8 domain walls,''
  Class.\ Quant.\ Grav.\  {\bf 18} (2001) 3359
 % doi:10.1088/0264-9381/18/17/303
  [hep-th/0103233].
  %%CITATION = doi:10.1088/0264-9381/18/17/303;%%
  %257 citations counted in INSPIRE as of 01 Dec 2019

\bibitem{Saltman}
  A.~Saltman and E.~Silverstein,
  %``A New handle on de Sitter compactifications,''
  JHEP {\bf 0601} (2006) 139
%  doi:10.1088/1126-6708/2006/01/139
  [hep-th/0411271].
  %%CITATION = doi:10.1088/1126-6708/2006/01/139;%%
  %93 citations counted in INSPIRE as of 01 Dec 2019

\bibitem{Craig1}
  N.~Craig, I.~Garcia Garcia and D.~Sutherland,
  %``Disassembling the Clockwork Mechanism,''
  JHEP {\bf 1710} (2017) 018
 % doi:10.1007/JHEP10(2017)018
  [arXiv:1704.07831 [hep-ph]].
  %%CITATION = doi:10.1007/JHEP10(2017)018;%%
  %40 citations counted in INSPIRE as of 01 Aug 2019
%\cite{Giudice:2017suc}
\bibitem{GM2}
  G.~F.~Giudice and M.~McCullough,
  %``Comment on "Disassembling the Clockwork Mechanism",''
  arXiv:1705.10162 [hep-ph].
  %%CITATION = ARXIV:1705.10162;%%
  %30 citations counted in INSPIRE as of 01 Aug 2019
  %\cite{Giudice:2017fmj}

%\cite{Randall:1999ee}
\bibitem{Randall:1999ee} 
  L.~Randall and R.~Sundrum,
  %``A Large mass hierarchy from a small extra dimension,''
  Phys.\ Rev.\ Lett.\  {\bf 83}, 3370 (1999)
  doi:10.1103/PhysRevLett.83.3370
  [hep-ph/9905221].
  %%CITATION = doi:10.1103/PhysRevLett.83.3370;%%
  %8379 citations counted in INSPIRE as of 05 Sep 2019
	
\bibitem{GM3}
  G.~F.~Giudice, Y.~Kats, M.~McCullough, R.~Torre and A.~Urbano,
  %``Clockwork/linear dilaton: structure and phenomenology,''
  JHEP {\bf 1806} (2018) 009
%  doi:10.1007/JHEP06(2018)009
  [arXiv:1711.08437 [hep-ph]].
  %%CITATION = doi:10.1007/JHEP06(2018)009;%%
  %30 citations counted in INSPIRE as of 01 Aug 2019
  %\cite{Teresi:2018eai}
\bibitem{Teresi}
  D.~Teresi,
  %``Clockwork without supersymmetry,''
  Phys.\ Lett.\ B {\bf 783} (2018) 1
  %doi:10.1016/j.physletb.2018.06, 10.1016/j.physletb.2018.06.049
  [arXiv:1802.01591 [hep-ph]].
  %%CITATION = doi:10.1016/j.physletb.2018.06, 10.1016/j.physletb.2018.06.049;%%
  %9 citations counted in INSPIRE as of 01 Aug 2019
%\cite{Ahmed:2016viu}
\bibitem{Ahmed}
  A.~Ahmed and B.~M.~Dillon,
  %``Clockwork Goldstone Bosons,''
  Phys.\ Rev.\ D {\bf 96} (2017) no.11,  115031
  doi:10.1103/PhysRevD.96.115031
  [arXiv:1612.04011 [hep-ph]].
  %%CITATION = doi:10.1103/PhysRevD.96.115031;%%
  %30 citations counted in INSPIRE as of 01 Aug 2019
%\cite{Hambye:2016qkf}
\bibitem{Hambye}
  T.~Hambye, D.~Teresi and M.~H.~G.~Tytgat,
  %``A Clockwork WIMP,''
  JHEP {\bf 1707} (2017) 047
%  doi:10.1007/JHEP07(2017)047
  [arXiv:1612.06411 [hep-ph]].
  %%CITATION = doi:10.1007/JHEP07(2017)047;%%
  %33 citations counted in INSPIRE as of 01 Aug 2019
  %\cite{Niedermann:2018lhx}
\bibitem{cwgravity}
  F.~Niedermann, A.~Padilla and P.~M.~Saffin,
  %``Higher Order Clockwork Gravity,''
  Phys.\ Rev.\ D {\bf 98} (2018) no.10,  104014
 % doi:10.1103/PhysRevD.98.104014
  [arXiv:1805.03523 [hep-th]].
  %%CITATION = doi:10.1103/PhysRevD.98.104014;%%
  %3 citations counted in INSPIRE as of 01 Aug 2019
  
  %\cite{Torre:2018jnf}
\bibitem{Torre}
  R.~Torre,
  %``Clockwork/Linear Dilaton: Structure and phenomenology,''
  arXiv:1806.04483 [hep-ph].
  %%CITATION = ARXIV:1806.04483;%%
  %1 citations counted in INSPIRE as of 01 Aug 2019
  %\cite{Agrawal:2018mkd}
\bibitem{Agrawal}
  P.~Agrawal, J.~Fan and M.~Reece,
  %``Clockwork Axions in Cosmology: Is Chromonatural Inflation Chrononatural?,''
  JHEP {\bf 1810} (2018) 193
%  doi:10.1007/JHEP10(2018)193
  [arXiv:1806.09621 [hep-th]].
  %%CITATION = doi:10.1007/JHEP10(2018)193;%%
  %17 citations counted in INSPIRE as of 01 Aug 2019
  %\cite{Goudelis:2018xqi}
\bibitem{fimp}
  A.~Goudelis, K.~A.~Mohan and D.~Sengupta,
  %``Clockworking FIMPs,''
  JHEP {\bf 1810} (2018) 014
 % doi:10.1007/JHEP10(2018)014
  [arXiv:1807.06642 [hep-ph]].
  %%CITATION = doi:10.1007/JHEP10(2018)014;%%
  %9 citations counted in INSPIRE as of 01 Aug 2019
  %\cite{Im:2018dum}
\bibitem{Im1}
  S.~H.~Im, H.~P.~Nilles and M.~Olechowski,
  %``Heterotic M-Theory from the Clockwork Perspective,''
  JHEP {\bf 1901} (2019) 151
 % doi:10.1007/JHEP01(2019)151
  [arXiv:1811.11838 [hep-th]].
  %%CITATION = doi:10.1007/JHEP01(2019)151;%%
  %2 citations counted in INSPIRE as of 01 Aug 2019
  %\cite{Im:2019cnl}
\bibitem{Im2}
  S.~H.~Im, H.~P.~Nilles and M.~Olechowski,
  %``Axion clockworks from heterotic M-theory: the QCD-axion and its ultra-light companion,''
  arXiv:1906.11851 [hep-th].
  %%CITATION = ARXIV:1906.11851;%%
  
  

%\cite{Enqvist:2007tb}
\bibitem{Martin}
  K.~Enqvist, S.~Hannestad and M.~S.~Sloth,
  %``Seesaw mechanism for scalar fields as possible basis for dark energy,''
  Phys.\ Rev.\ Lett.\  {\bf 99} (2007) 031301
%  doi:10.1103/PhysRevLett.99.031301
  [hep-ph/0702236 [HEP-PH]].
  %%CITATION = doi:10.1103/PhysRevLett.99.031301;%%
  %7 citations counted in INSPIRE as of 01 Aug 2019
  
  
  %\cite{Graham:2015cka}
\bibitem{relaxion}
  P.~W.~Graham, D.~E.~Kaplan and S.~Rajendran,
  %``Cosmological Relaxation of the Electroweak Scale,''
  Phys.\ Rev.\ Lett.\  {\bf 115} (2015) no.22,  221801
 % doi:10.1103/PhysRevLett.115.221801
  [arXiv:1504.07551 [hep-ph]].
  %%CITATION = doi:10.1103/PhysRevLett.115.221801;%%
  %299 citations counted in INSPIRE as of 01 Aug 2019

  
  
  %\cite{Dvali:2004tma}
\bibitem{Dvali1}
  G.~Dvali,
  %``Large hierarchies from attractor vacua,''
  Phys.\ Rev.\ D {\bf 74} (2006) 025018
 % doi:10.1103/PhysRevD.74.025018
  [hep-th/0410286].
  %%CITATION = doi:10.1103/PhysRevD.74.025018;%%
  %64 citations counted in INSPIRE as of 02 Aug 2019
%\cite{Dvali:2005an}
\bibitem{Dvali2}
  G.~Dvali,
  %``Three-form gauging of axion symmetries and gravity,''
  hep-th/0507215.
  %%CITATION = HEP-TH/0507215;%%
  %93 citations counted in INSPIRE as of 02 Aug 2019
  
  %\cite{Kaloper:2008fb}
\bibitem{KS1}
  N.~Kaloper and L.~Sorbo,
  %``A Natural Framework for Chaotic Inflation,''
  Phys.\ Rev.\ Lett.\  {\bf 102} (2009) 121301
%  doi:10.1103/PhysRevLett.102.121301
  [arXiv:0811.1989 [hep-th]].
  %%CITATION = doi:10.1103/PhysRevLett.102.121301;%%
  %260 citations counted in INSPIRE as of 02 Aug 2019
  %\cite{Kaloper:2008qs}
\bibitem{KS2}
  N.~Kaloper and L.~Sorbo,
  %``Where in the String Landscape is Quintessence,''
  Phys.\ Rev.\ D {\bf 79} (2009) 043528
 % doi:10.1103/PhysRevD.79.043528
  [arXiv:0810.5346 [hep-th]].
  %%CITATION = doi:10.1103/PhysRevD.79.043528;%%
  %45 citations counted in INSPIRE as of 02 Aug 2019
  %\cite{Kaloper:2011jz}
\bibitem{KLS}
  N.~Kaloper, A.~Lawrence and L.~Sorbo,
  %``An Ignoble Approach to Large Field Inflation,''
  JCAP {\bf 1103} (2011) 023
%  doi:10.1088/1475-7516/2011/03/023
  [arXiv:1101.0026 [hep-th]].
  %%CITATION = doi:10.1088/1475-7516/2011/03/023;%%
  %235 citations counted in INSPIRE as of 02 Aug 2019
  
  
  %\cite{Silverstein:2008sg}
\bibitem{monod1}
  E.~Silverstein and A.~Westphal,
  %``Monodromy in the CMB: Gravity Waves and String Inflation,''
  Phys.\ Rev.\ D {\bf 78} (2008) 106003
%  doi:10.1103/PhysRevD.78.106003
  [arXiv:0803.3085 [hep-th]].
  %%CITATION = doi:10.1103/PhysRevD.78.106003;%%
  %656 citations counted in INSPIRE as of 12 Aug 2019
  %\cite{McAllister:2008hb}
\bibitem{monod2}
  L.~McAllister, E.~Silverstein and A.~Westphal,
  %``Gravity Waves and Linear Inflation from Axion Monodromy,''
  Phys.\ Rev.\ D {\bf 82} (2010) 046003
%  doi:10.1103/PhysRevD.82.046003
  [arXiv:0808.0706 [hep-th]].
  %%CITATION = doi:10.1103/PhysRevD.82.046003;%%
  %598 citations counted in INSPIRE as of 12 Aug 2019
  
  %\cite{Kaloper:2016fbr}
\bibitem{london}
  N.~Kaloper and A.~Lawrence,
  %``London equation for monodromy inflation,''
  Phys.\ Rev.\ D {\bf 95} (2017) no.6,  063526
 % doi:10.1103/PhysRevD.95.063526
  [arXiv:1607.06105 [hep-th]].
  %%CITATION = doi:10.1103/PhysRevD.95.063526;%%
  %22 citations counted in INSPIRE as of 12 Aug 2019
  
  %\cite{Padilla:2018hvp}
\bibitem{monap}
  A.~Padilla,
  %``Monodromy inflation and an emergent mechanism for stabilising the cosmological constant,''
  JHEP {\bf 1901} (2019) 175
%  doi:10.1007/JHEP01(2019)175
  [arXiv:1806.04740 [hep-th]].
  %%CITATION = doi:10.1007/JHEP01(2019)175;%%
  %3 citations counted in INSPIRE as of 12 Aug 2019
%\cite{El-Menoufi:2019qva}
\bibitem{monbasem}
  B.~K.~El-Menoufi, S.~Nagy, F.~Niedermann and A.~Padilla,
  %``Quantum corrections to vacuum energy sequestering (with monodromy),''
  arXiv:1903.07612 [hep-th].
  %%CITATION = ARXIV:1903.07612;%%
  %\cite{Kaloper:2018kma}
\bibitem{monnk}
  N.~Kaloper,
  %``Irrational Monodromies of Vacuum Energy,''
  arXiv:1806.03308 [hep-th].
  %%CITATION = ARXIV:1806.03308;%%
  %4 citations counted in INSPIRE as of 12 Aug 2019
  
  %\cite{Kaloper:2013zca}
\bibitem{KP1}
  N.~Kaloper and A.~Padilla,
  %``Sequestering the Standard Model Vacuum Energy,''
  Phys.\ Rev.\ Lett.\  {\bf 112} (2014) 9,  091304.
 % [arXiv:1309.6562 [hep-th]].
  %%CITATION = ARXIV:1309.6562;%%
  %22 citations counted in INSPIRE as of 17 mar 2015
  
  
  %\cite{Kaloper:2014dqa}
\bibitem{KP2}
  N.~Kaloper and A.~Padilla,
  %``Vacuum Energy Sequestering: The Framework and Its Cosmological Consequences,''
  Phys.\ Rev.\ D {\bf 90} (2014) 8,  084023
   [Addendum-ibid.\ D {\bf 90} (2014) 10,  109901].
%  [arXiv:1406.0711 [hep-th]].
  %%CITATION = ARXIV:1406.0711;%%
  %10 citations counted in INSPIRE as of 17 Mar 2015
  
  %\cite{Kaloper:2014fca}
\bibitem{KP3}
 N.~Kaloper and A.~Padilla,
  %``Sequestration of Vacuum Energy and the End of the Universe,''
  Phys.\ Rev.\ Lett.\  {\bf 114} (2015) 10,  101302
 % doi:10.1103/PhysRevLett.114.101302
  %[arXiv:1409.7073 [hep-th]].
  %%CITATION = doi:10.1103/PhysRevLett.114.101302;%%
  %12 citations counted in INSPIRE as of 06 May 2016
  
\bibitem{KPSZ} 
  N.~Kaloper, A.~Padilla, D.~Stefanyszyn and G.~Zahariade,
  %``Manifestly Local Theory of Vacuum Energy Sequestering,''
  Phys.\ Rev.\ Lett.\  {\bf 116} (2016) 5,  051302
%  doi:10.1103/PhysRevLett.116.051302
  %[arXiv:1505.01492 [hep-th]].
  %%CITATION = doi:10.1103/PhysRevLett.116.051302;%%
  %9 citations counted in INSPIRE as of 06 May 2016
  
  %\cite{Kaloper:2016yfa}
\bibitem{KPS}
  N.~Kaloper, A.~Padilla and D.~Stefanyszyn,
  %``Sequestering effects on and of vacuum decay,''
  Phys.\ Rev.\ D {\bf 94} (2016) no.2,  025022
%  doi:10.1103/PhysRevD.94.025022
  [arXiv:1604.04000 [hep-th]].
  %%CITATION = doi:10.1103/PhysRevD.94.025022;%%
  %5 citations counted in INSPIRE as of 11 Jun 2018
  
  %\cite{Kaloper:2016jsd}
\bibitem{KP4}
  N.~Kaloper and A.~Padilla,
  %``Vacuum Energy Sequestering and Graviton Loops,''
  Phys.\ Rev.\ Lett.\  {\bf 118} (2017) no.6,  061303
  %doi:10.1103/PhysRevLett.118.061303
  [arXiv:1606.04958 [hep-th]].
  %%CITATION = doi:10.1103/PhysRevLett.118.061303;%%
  %4 citations counted in INSPIRE as of 12 May 2017
  
  %\cite{DAmico:2017ngr}
\bibitem{etude}
  G.~D'Amico, N.~Kaloper, A.~Padilla, D.~Stefanyszyn, A.~Westphal and G.~Zahariade,
  %``An �tude on global vacuum energy sequester,''
  JHEP {\bf 1709} (2017) 074
%  doi:10.1007/JHEP09(2017)074
  [arXiv:1705.08950 [hep-th]].
  %%CITATION = doi:10.1007/JHEP09(2017)074;%%
  %7 citations counted in INSPIRE as of 11 Jun 2018
  
  %\cite{Polchinski:1995sm}
\bibitem{Pol1}
  J.~Polchinski and A.~Strominger,
  %``New vacua for type II string theory,''
  Phys.\ Lett.\ B {\bf 388} (1996) 736
%  doi:10.1016/S0370-2693(96)01219-1
  [hep-th/9510227].
  %%CITATION = doi:10.1016/S0370-2693(96)01219-1;%%
  %343 citations counted in INSPIRE as of 12 Aug 2019
%\cite{Bousso:2000xa}
\bibitem{BP}
  R.~Bousso and J.~Polchinski,
  %``Quantization of four form fluxes and dynamical neutralization of the cosmological constant,''
  JHEP {\bf 0006} (2000) 006
 % doi:10.1088/1126-6708/2000/06/006
  [hep-th/0004134].
  %%CITATION = doi:10.1088/1126-6708/2000/06/006;%%
  %1082 citations counted in INSPIRE as of 12 Aug 2019
  
  
  %\cite{Lee:1987qc}
\bibitem{dsds}
K.~Lee and E.~J.~Weinberg,
%``Decay of the True Vacuum in Curved Space-time,''
Phys.\ Rev.\ D \textbf{36} (1987), 1088
%doi:10.1103/PhysRevD.36.1088
%138 citations counted in INSPIRE as of 27 Mar 2020  


  
  
  \bibitem{future}
  L.~Bordin, F.~Cunillera,  A. ~Leh\'ebel and A. ~Padilla,
Work in progress.

  
\end{thebibliography}
\end{document}